\documentclass[conference]{IEEEtran}
\IEEEoverridecommandlockouts

\usepackage[utf8]{inputenc}
\usepackage[T1]{fontenc}

\usepackage{CJKutf8}
\usepackage{datetime}
\usepackage{amsmath}
\usepackage{amssymb}
\usepackage{mathtools}
\usepackage{subcaption}
\usepackage{xcolor}
\usepackage{ulem}
\usepackage{hyperref}
\usepackage{graphicx}
\usepackage{cite}
\usepackage{multirow}
\usepackage{makecell}
\usepackage{url}
\usepackage{placeins}

\graphicspath{{images/}}
\DeclareGraphicsExtensions{.pdf}

\begin{document}

\title{Ensuring Fairness with Transparent Auditing of Quantitative Bias in AI Systems\\
\thanks{This project is partially supported by MOST111-2221-E-001-010-MY2 and 112-2634-F-002-004 from National Science and Technology Council and AS-GCS-113-H04 from Academia Sinica.}
}

\DeclarePairedDelimiter{\set}{\{}{\}}
\DeclarePairedDelimiter{\tuple}{(}{)}
\DeclarePairedDelimiter{\abs}{\lvert}{\rvert}

\renewcommand{\implies}{\rightarrow}

\newcommand{\db}{\mathcal{D}}
\newcommand{\model}{\mathcal{M}}
\newcommand{\priv}{R}
\newcommand{\pos}{\hat{P}}
\newcommand{\posm}{\hat{P}_\mathcal{M}}
\newcommand{\tru}{T}
\newcommand{\leg}{L}
\newcommand{\sco}{\hat{S}}
\newcommand{\scom}{\hat{S}_\mathcal{M}}
\newcommand{\calib}{C}

\newcommand{\bye}[1]{}

\author{\IEEEauthorblockN{Chih-Cheng Rex Yuan}
\IEEEauthorblockA{\textit{Institute of Information Science} \\
\textit{Academia Sinica}\\
Taipei, Taiwan \\
hello@rexyuan.com}
\and
\IEEEauthorblockN{Bow-Yaw Wang}
\IEEEauthorblockA{\textit{Institute of Information Science} \\
\textit{Academia Sinica}\\
Taipei, Taiwan \\
bywang@iis.sinica.edu.tw}
}

\maketitle

\begin{abstract}
With the rapid advancement of AI, there is a growing trend to integrate AI into decision-making processes. However, AI systems may exhibit biases that lead decision-makers to draw unfair conclusions. Notably, the COMPAS system used in the American justice system to evaluate recidivism was found to favor racial majority groups; specifically, it violates a fairness standard called equalized odds. Various measures have been proposed to assess AI fairness. We present a framework for auditing AI fairness, involving third-party auditors and AI system providers, and we have created a tool to facilitate systematic examination of AI systems. The tool is open-sourced and publicly available. Unlike traditional AI systems, we advocate a transparent white-box and statistics-based approach. It can be utilized by third-party auditors, AI developers, or the general public for reference when judging the fairness criterion of AI systems.
\end{abstract}


\section{Introduction}
The accelerating pace of artificial intelligence (AI) technologies has revolutionized numerous fields in recent years. In healthcare, AI-driven diagnostic systems streamline disease identification, while in finance, automated trading algorithms analyze market trends to execute optimal trades swiftly. This remarkable progress has reached diverse industries, sprouting out various applications for different purposes.

One of its most profound impacts is how AI has transformed decision-making processes across sectors. By harnessing vast amounts of data and employing advanced algorithms, AI empowers organizations to make more informed and strategic decisions. From assessing job applicants to determining school admissions, AI-driven insights offer efficient and analytical advantages.

AI systems, however, may be biased. Factors such as inherent biases in original data sets or flaws in algorithm designs could contribute to bias within AI systems. When left unchecked, the ramifications of such biases extend far beyond mere inaccuracies; they can lead to devastating consequences, such as amplifying systemic injustices, perpetuating group discrimination, and exacerbating societal inequalities.

Correctional Offender Management Profiling for Alternative Sanctions (COMPAS) is a tool for predicting recidivism---the tendency of criminals to reoffend. It is used in the criminal justice system in multiple states in the United States. In 2016, it was discovered in an investigation by the journalists at ProPublica that the COMPAS system is, in fact, unfair towards minority and disadvantaged groups.

Cases such as COMPAS underscore the critical importance of rigorously examining the fairness of AI systems by third parties. Fairness is fundamentally a subjective social construct, heavily influenced by cultural context and deeply rooted in historical inequalities. However, there have been developments in research leveraging statistical metrics to quantify fairness that provide transparent and objective insights.

These statistical metrics offer a systematic approach to evaluating fairness across various dimensions of AI systems. Among these metrics are disparate impact and demographic parity, to name a few. In particular, the findings of the COMPAS report revealed that it violates a critical metric known as equalized odds.

The function of our framework is to make comprehensive reviewing of statistical metrics accessible to third party auditors. Central to this role is the precise definition and flexible abstraction of the metrics. For example, our framework would reveal that the COMPAS dataset not only violates equalized odds but also conditional statistical parity and mean difference.

Including independent and trusted auditors is pivotal for objectivity and accountability. Moreover, the involvement of third parties is often indispensable due to their specialized expertise in relevant domains. Auditors, equipped with our tool, can thoroughly review AI systems for bias and fairness violations.

Our tool builds upon transparent definitions widely accepted within the scientific community. By providing an abstraction layer atop these definitions, we enhance their flexibility and transparency, which enables users to adapt them to diverse contexts with ease and reveal them to the public. Moreover, we prioritize transparency throughout our tool's development process, offering access to the source code for scrutiny, thereby ensuring transparency from conception to implementation.

Our open-sourced tool is written in Python and is offered as a Python package. It supports common dataset formats such as CSV. The package is available at \url{https://pypi.org/project/fairness-checker/}.

\section{Fairness Measures}
\label{sec:measures}
Fairness is about making sure the disadvantaged and unprivileged groups of individuals are treated equitably. However, there have been several interpretations of it proposed in the past. To have a constructive discussion on fairness, we must first have precise definitions of it. Pessach and Shmueli \cite{pessach2022review} formulated a number of fairness measures in a unified mathematical notation. We will base our framework on their formulation.

\subsection{Preliminaries}
For the following definition, we will use $Y$ to denote the ground truth of an outcome; $\hat{Y}$ to denote the predicated result of an outcome; $Y = 1$ and $\hat{Y} = 1$ to denote them being accepted or positive. For example, let $Y$ be recidivism. Then $Y = 1$ means the case of an individual actually recidivating and $\hat{Y} = 0$ means the prediction of an individual recidivating is negative. In addition, we use $V$ and $\hat{V}$ when the truth and the prediction are not binary. For example, we denote the COMPAS score by $\hat{V}$, which ranges from 1 to 10.

\emph{Protected attributes} are the characteristics of individuals that are, for example, legally or ethically, considered sensitive and warrant protection against discrimination and bias. We denote by $S$ some protected attribute. We write $S = 1$ to represent the \emph{privileged group} and $S \neq 1$ to represent the \emph{unprivileged group}. For example, let $S$ be Caucasian. Then $S = 1$ represents the case of an individual's race being Caucasian and $S \neq 1$ vice versa.

We denote by $\epsilon$ some threshold that is used to limit the fairness measures.

\subsection{Disparate Impact}
In 1971 \cite{griggs1971}, the US Supreme Court ruled that it is illegal for hiring decisions to have  ``disparate impact'' by race, thus coining the term. It is taken as unintentional discrimination, as opposed to intentional discrimination, which is called ``disparate treatment''.

Legal cases involving disparate impact often refer to the ``80\% Rule'', advocated by the US Equal Employment Opportunity Commission \cite{eeoc1979}, where it requires the selection rate of a minority group to be no less than 80\% of that of a majority group. Formally \cite{feldman2015certifying}, it is:
\[
    \frac{P[\hat{Y} = 1 | S \neq 1]}{P[\hat{Y} = 1 | S = 1]} \geq 1 - \epsilon
\]
In the case of 80\% rule, $\epsilon = 20\%$.

\subsection{Demographic Parity}
Demographic parity \cite{pessach2022review}, also known as statistical parity, is similar to disparate impact but, instead of ratio, the difference is taken. It is named so to suggest that each demographic group(such as race, gender, or age) should have equal representation or opportunity. Formally, it is:
\[
    \abs{P[\hat{Y} = 1 | S = 1] - P[\hat{Y} = 1 | S \neq 1]} \leq \epsilon
\]

\subsection{Conditional Statistical Parity}
Conditional statistical parity \cite{corbett2017algorithmic} is similar to demographic parity, but, in addition to protected attributes, it further takes into account some ``legitimate'' attributes that are \textit{legitimately} related to the case. For example, a legitimate attribute when considering future recidivism could be the number of prior crimes committed. Formally, it is:
\[
    \abs{P[\hat{Y} = 1 | S = 1, L = l] - P[\hat{Y} = 1 | S \neq 1, L = l]} \leq \epsilon
\]
where $L$ denotes the legitimate attributes.

\subsection{Overall Accuracy Equality}
Overall accuracy equality \cite{berk2021fairness} is similar to demographic parity, but instead of the case of $\hat{Y} = 1$, it considers the case of $Y = \hat{Y}$; that is, the case where the prediction is accurate. Formally, it is:
\[
    \abs{P[Y = \hat{Y} | S = 1] - P[Y = \hat{Y} | S \neq 1]} \leq \epsilon
\]

\subsection{Mean Difference}
Mean difference \cite{vzliobaite2017measuring} considers the expected value of the prediction. Formally, it is:
\[
    \abs{E[\hat{Y}|S = 1] - E[\hat{Y}|S \neq 1]} \leq \epsilon
\]

\subsection{Equalized Odds}
Equalized odds \cite{hardt2016equality} is similar to demographic parity, but it further takes into account the ground truth. It considers the cases of true positive and false positive. It solves the downsides of a fully accurate classifier that might be deemed unfair by measures that do not consider ground truth. For example, consider a group $A$ that is predicated to recidivate and they do in fact recidivate and a group $B$ that is predicated to recidivate but end up never recidivating. A fully accurate classifier will always predict $A$ to recidivate while $B$ to never recidivate, and this will violate demographic parity. By taking ground truth into account, equalized odds avoids these pitfalls. Formally, it is:
\begin{align*}
    \abs{P[\hat{Y} = 1 | S = 1, Y = 0] - P[\hat{Y} = 1 | S \neq 1, Y = 0]} & \leq \epsilon \\
    \abs{P[\hat{Y} = 1 | S = 1, Y = 1] - P[\hat{Y} = 1 | S \neq 1, Y = 1]} & \leq \epsilon
\end{align*}

\subsection{Equal Opportunity}
Equal opportunity \cite{hardt2016equality} is a relaxation of equalized odds by only considering the true positive case. Formally, it is:
\begin{align*}
    \abs{P[\hat{Y} = 1 | S = 1, Y = 1] - P[\hat{Y} = 1 | S \neq 1, Y = 1]} & \leq \epsilon
\end{align*}

\subsection{Predictive Equality}
Predictive equality \cite{corbett2017algorithmic} is also a relaxation of equalized odds by only considering the false positive case. Formally, it is:
\begin{align*}
    \abs{P[\hat{Y} = 1 | S = 1, Y = 0] - P[\hat{Y} = 1 | S \neq 1, Y = 0]} \leq \epsilon
\end{align*}

\subsection{Conditional Use Accuracy Equality}
Conditional use accuracy equality \cite{berk2021fairness} is similar to equalized odds, but instead of conditioning on the ground truth, it conditions on the prediction and calculates the probability of the ground truth. It can be seen as checking the prediction accuracy across groups, thus the name. It further requires the measure in the case of positive predictive values to be less than that of the negative predictive values. Formally, it is:
\begin{align*}
    \abs{P[Y = 1 | S = 1, \hat{Y} = 1] & - P[Y = 1 | S \neq 1, \hat{Y} = 1]} && \leq \epsilon \\
    & \rotatebox{90}{>} && \\
    \abs{P[Y = 0 | S = 1, \hat{Y} = 0] & - P[Y = 0 | S \neq 1, \hat{Y} = 0]} && \leq \epsilon
\end{align*}

\subsection{Predictive Parity}
Predictive parity \cite{chouldechova2017fair} is a relaxation of conditional use accuracy equality by only considering the positive predictive value case. Formally, it is:
\begin{align*}
    \abs{P[Y = 1 | S = 1, \hat{Y} = 1] - P[Y = 1 | S \neq 1, \hat{Y} = 1]} \leq \epsilon
\end{align*}

\subsection{Equal Calibration}
Equal calibration \cite{chouldechova2017fair} is similar to equal opportunity, but instead of having a binary $\hat{Y}$, it is conditioned on the range of the predicted value $\hat{V}$. For example, this could be conditioned on the highest COMPAS score $\hat{V} = 10$. Calibration is a concept of having a fair score function \cite{fraenkel2020fairness}. Formally, it is:
\begin{align*}
    \abs{P[Y = 1 | S = 1, \hat{V} = v] - P[Y = 1 | S \neq 1, \hat{V} = v]} \leq \epsilon
\end{align*}

\subsection{Positive Balance}
Positive balance \cite{kleinberg2016inherent} is similar to equal opportunity, but instead of taking the difference of probability of binary prediction $\hat{Y}$, it takes the difference of the expected value of the score $\hat{V}$, which may be non-binary, such as the score of COMPAS. Formally, it is:
\begin{align*}
    \abs{E[\hat{V} | Y = 1, S = 1] - E[\hat{V} | Y = 1, S \neq 1]} \leq \epsilon
\end{align*}

\subsection{Negative Balance}
Negative balance \cite{kleinberg2016inherent} is like positive balance except it conditions on the case of $Y = 0$. Formally, it is:
\begin{align*}
    \abs{E[\hat{V} | Y = 0, S = 1] - E[\hat{V} | Y = 0, S \neq 1]} \leq \epsilon
\end{align*}

These fairness measures are compiled in Table~\ref{tab:measures}.

\begin{table*}[h]
    \centering
    \begin{tabular}{|l|c|} 
        \hline
        \textbf{Fairness Measure} & \textbf{Definition} \\
        \hline
Disparate Impact & $\frac{P[\hat{Y} = 1 | S \neq 1]}{P[\hat{Y} = 1 | S = 1]} \geq 1 - \epsilon$ \\
\hline
Demographic Parity & $\abs{P[\hat{Y} = 1 | S = 1] - P[\hat{Y} = 1 | S \neq 1]} \leq \epsilon$ \\
\hline
Conditional Statistical Parity & $\abs{P[\hat{Y} = 1 | S = 1, L = l] - P[\hat{Y} = 1 | S \neq 1, L = l]} \leq \epsilon$ \\
\hline
Overall Accuracy Equality & $\abs{P[Y = \hat{Y} | S = 1] - P[Y = \hat{Y} | S \neq 1]} \leq \epsilon$ \\
\hline
Mean Difference & $\abs{E[\hat{Y}|S = 1] - E[\hat{Y}|S \neq 1]} \leq \epsilon$ \\
\hline
\multirow{2}{*}{Equalized Odds} & $\abs{P[\hat{Y} = 1 | S = 1, Y = 0] - P[\hat{Y} = 1 | S \neq 1, Y = 0]} \leq \epsilon$ \\
\cline{2-2}
                                & $\abs{P[\hat{Y} = 1 | S = 1, Y = 1] - P[\hat{Y} = 1 | S \neq 1, Y = 1]} \leq \epsilon$ \\
\hline
Equal Opportunity & $\abs{P[\hat{Y} = 1 | S = 1, Y = 1] - P[\hat{Y} = 1 | S \neq 1, Y = 1]} \leq \epsilon$ \\
\hline
Predictive Equality & $\abs{P[\hat{Y} = 1 | S = 1, Y = 0] - P[\hat{Y} = 1 | S \neq 1, Y = 0]} \leq \epsilon$ \\
\hline
\multirow{2}{*}{Conditional Use Accuracy Equality} & $\abs{P[Y = 1 | S = 1, \hat{Y} = 1] - P[Y = 1 | S \neq 1, \hat{Y} = 1]} \leq \epsilon$ \\
\cline{2-2}
                                 & $\abs{P[Y = 0 | S = 1, \hat{Y} = 0] - P[Y = 0 | S \neq 1, \hat{Y} = 0]} \leq \epsilon$ \\
\hline
Predictive Parity & $\abs{P[Y = 1 | S = 1, \hat{Y} = 1] - P[Y = 1 | S \neq 1, \hat{Y} = 1]} \leq \epsilon$ \\
\hline
Equal Calibration & $\abs{P[Y = 1 | S = 1, \hat{V} = v] - P[Y = 1 | S \neq 1, \hat{V} = v]} \leq \epsilon$ \\
\hline
Positive Balance & $\abs{E[\hat{V} | Y = 1, S = 1] - E[\hat{V} | Y = 1, S \neq 1]} \leq \epsilon$ \\
\hline
Negative Balance & $\abs{E[\hat{V} | Y = 0, S = 1] - E[\hat{V} | Y = 0, S \neq 1]} \leq \epsilon$ \\
\hline
    \end{tabular}
    \caption{Fairness measures.}
    \label{tab:measures}
\end{table*}

\section{Auditing Framework}
\label{sec:framework}
Per the review by Pessach and Shmueli \cite{pessach2022review}, we designed an auditing framework for calculating the various fairness measures. We offer two versions of fairness checkers: one for when the prediction results are readily available in CSV input and one for when a model is provided. In most auditing cases the model version is preferred because CSV results can be easily fabricated.

\subsection{Notations}
Types $\alpha,\beta$ are stand-ins for any type. Let $\alpha \implies \beta$ denote a function from type $\alpha$ to type $\beta$. $\alpha_i$ denotes some particular type; $\prod_i \alpha_i$ denotes the Cartesian product of multiple $\alpha$ of possibly different types. We write $x : \alpha$ to mean $x$ is of the type $\alpha$. We write $\mathtt{str}$ for the string type, $\mathtt{bool}$ for the boolean type, and $\mathtt{int}$ for the integer type.

A \emph{database} $\db = \set{r_1, r_2, ...}$ is a collection of rows. A \emph{row} $r_i : \mathtt{key} \implies \mathtt{value}$ is a lookup table or dictionary, where the concrete type of $\mathtt{key}$ and $\mathtt{value}$ is $\mathtt{str}$. For example, $r_n(\text{``sex''}) = \text{``Female''}$ means $r_n$'s sex is female. Henceforth, we will use $\mathtt{row}$ as a type synonym of $\mathtt{key} \implies \mathtt{value}$.

A \emph{model} $\model : \mathtt{row} \implies \alpha$ is a black-box predictor that takes a row and returns its prediction result of some type $\alpha$; for example, if the model returns a string then $\alpha = \mathtt{str}$.

A \emph{privileged predicate} $\priv : \mathtt{row} \implies \mathtt{bool}$ takes a row and determines if it belongs to the privileged group. For example, $\priv(r_i) := r_i(\text{``race''}) == \text{``Caucasian''}$ means the privileged group is those with race being Caucasian.

A \emph{positive predicate} of rows $\pos : \mathtt{row} \implies \mathtt{bool}$ takes a row and determines if its prediction is positive. For example, $\pos(r_i) := \mathtt{int}(r_i(\text{``score''})) > 7$ means a row's prediction is positive if its score is greater than 7. A positive predicate of model results $\posm : \alpha \implies \mathtt{bool}$ takes a model's result and determines if it is positive. In the simplest case, the model returns a $\mathtt{bool}$, and the positive predicate can just be the identity function.

A \emph{score predicate} of rows $\sco : \mathtt{row} \implies \beta$ takes a row and gives its predicted score of some type $\beta$. Similar to positive predicate, a score predicate of model results $\scom : \alpha \implies \beta$ takes a model's result and gives its score.

A \emph{ground truth predicate} $\tru : \mathtt{row} \implies \mathtt{bool}$ takes a row and gives the ground truth of the result.

A \emph{legitimate predicate} $\leg : \prod_i \alpha_i \implies \mathtt{row} \implies \mathtt{bool}$ takes $n$ parameters and returns a row predicate. For example, $\leg(x)(r_i) := \mathtt{int}(r_i(\text{``priors\_count''})) > x$ first takes a parameter $x$ and then decides if priors count is larger than it.

A \emph{calibration predicate} $\calib : \prod_i \alpha_i \implies \mathtt{row} \implies \mathtt{bool}$, takes $n$ parameters and returns a row predicate. For example, $\calib(u, l)(r_i) := l < \mathtt{int}(r_i(\text{``score''})) < u$ first takes two parameters $u,l$ as upper bound and lower bound, and then decides if $r_i$'s prediction score is between them.

\subsection{Definitions}
We abstracted the idea of privileged groups and positive prediction as predicates to maximize flexibility. With proper predicates, any model output can be used; even prose-like responses of generative models can be included given adequate predicates.

Given a dataset $\db$ for auditing use, and given CSV prediction results or the prediction model $\model$ itself, we can calculate the fairness measures by modeling the statistical random variables mentioned in Section~\ref{sec:measures} with our predicates. We demonstrate this process in the following for a few selected measures; all the other measures can be modeled similarly.

For equal opportunity, recall its formal definition:
\begin{align*}
    \abs{P[\hat{Y} = 1 | S = 1, Y = 1] - P[\hat{Y} = 1 | S \neq 1, Y = 1]} \leq \epsilon
\end{align*}
To model $Y$, $\hat{Y}$, and $S$, we define the corresponding $\tru$, $\pos$, and $\priv$. We have $Y = 1$ if and only if $\tru$ is true; $\hat{Y} = 1$ if and only if $\pos$ is true; $S = 1$ if and only if $\priv$ is true; and vice versa. This way, we can calculate equal opportunity as a function:
\begin{align*}
    \text{equal\_opportunity} (\epsilon, \model, \priv, \pos, \tru)
\end{align*}

For positive balance, its formal definition is:
\begin{align*}
    \abs{E[\hat{V} | Y = 0, S = 1] - E[\hat{V} | Y = 0, S \neq 1]} \leq \epsilon
\end{align*}
We model $Y$ and $S$ as the previous example. As for $\hat{V}$ we have it equal to the output of $\sco$, and now we can calculate the measure. We calculate positive balance by calling the function:
\begin{align*}
    \text{positive\_balance} (\epsilon, \model, \priv, \sco, \tru)
\end{align*}

For equal calibration, recall its formal definition:
\begin{align*}
    \abs{P[Y = 1 | S = 1, \hat{V} = v] - P[Y = 1 | S \neq 1, \hat{V} = v]} \leq \epsilon
\end{align*}
Here we model $Y$ and $S$ as above. As for $\hat{V} = v$, we abstracted $\hat{V}$ with $\calib$ and $v$ with $args$. We have $\hat{V} = v$ if and only if $\calib(v)(r_i)$ is true. We calculate equal calibration by calling the function:
\begin{align*}
    \text{equal\_calibration} (\epsilon, \model, \priv, \calib, \tru, (v))
\end{align*}

For conditional statistical parity, its formal definition is:
\[
    \abs{P[\hat{Y} = 1 | S = 1, L = l] - P[\hat{Y} = 1 | S \neq 1, L = l]} \leq \epsilon
\]
$\hat{Y}$ and $S$ are modeled as above. As for $L = l$, we abstracted it similarly to the case above by having $L = l$ if and only if $\leg(l)(r_i)$ is true. We calculate conditional statistical parity by calling the function:
\begin{align*}
    \text{conditional\_statistical\_parity} (\epsilon, \model, \priv, \pos, \leg, (l))
\end{align*}

This way, we can calculate each condition programmatically by testing if the predicate is true on all rows and obtain the resulting measure. All the measures can be modeled in a similar vein.

By calculating all the available fairness measures, we provide the auditors a comprehensive perspective on the fairness performance of a model or dataset.

We implemented the framework in Python, and it is available as a public domain open-sourced Python package at \url{https://pypi.org/project/fairness-checker/}.

\section{Application}
\subsection{Setup}
\label{sec:app}
In this section, we will apply the proposed framework to the ProPublica COMPAS dataset \cite{angwin2016machine,larson2016compas,larson2016propublica}.

Correctional Offender Management Profiling for Alternative Sanctions (COMPAS), developed by the private company Northpointe (now Equivant), is a risk assessment software used in the American criminal justice system to evaluate the likelihood of a defendant reoffending. Defendants taking the COMPAS test are given a questionnaire about topics ranging from family history to personal ideology. The questionnaire is then fed to the software system along with a number of parameters like the defendants' age, and then the system will assign a risk score to them from 1-10, with 10 being the highest risk.

ProPublica is an American non-profit journalism organization focused on public interests. In 2016, they conducted an investigative report into the COMPAS system. They obtained the 2013-2014 COMPAS score data of over 10,000 defendants in Florida. They also obtained criminal records of these defendants through 2016 and compared if they actually recidivate or not. They only counted misdemeanors and felonies as recidivism but not less serious crimes such as infractions. In the study, they have found that black defendants are disproportionately scored higher than they actually are, and white defendants are disproportionately scored lower than they actually are.

We shall start applying our framework. There is generally no formal guide on how to set $\epsilon$, so we will take the 80\% rule's case and set it as $\epsilon = 0.2$.

We will set the unprivileged predicate to be ``African-American'' so that the unprivileged group will be African-American. Then we test if the African-American race is discriminated.
\[
    \priv(r_i) := r_i(\text{``race''}) \neq \text{``African-American''}
\]

As ProPublica referenced Northpointe's COMPAS Practitioners Guide and cited that ``medium''(5-7) and ``high''(8-10) categories of scores are considered to indicate a risk of recidivism, we set the positive predicate using the readily available category.
\[
    \pos(r_i) := r_i(\text{``score\_text''}) \in \set{\text{``Medium''},\text{``High''}}
\]

For the ground truth predicate, the recidivism data is already present in the ProPublica dataset, so all we have to do is a simple lookup.
\[
    \tru(r_i) := r_i(\text{``two\_year\_recid''}) == \text{``1''}
\]

For the score predicate, it is again a simple lookup.
\[
    \sco(r_i) := \mathtt{int}(r_i(\text{``decile\_score''}))
\]

For the legitimate predicate, we may want to look at defendants with priors.
\[
    \leg(0)(r_i) := \mathtt{int}(r_i(\text{``priors\_count''})) > 0
\]

For the calibration predicate, we may want to look at risk scores within a specific range.
\[
    \calib(7,5)(r_i) := 5 \leq \mathtt{int}(r_i(\text{``decile\_score''})) \leq 7
\]

With these predicates defined, we can call the fairness measure functions and check if the measures hold or not. The results are compiled in Table~\ref{tab:results}. Since some fairness measures are equivalent, we write them in the same entry.

\begin{table}[h]
    \centering
    \begin{tabular}{|l|c|c|} 
        \hline
        \textbf{Fairness Measure} & \textbf{Criterion} & \textbf{Pass} \\
        \hline
Disparate Impact & 1.81 > 0.8 & YES \\
\hline
Demographic Parity & 0.26 < 0.2 & NO \\
\hline
Conditional Statistical Parity & 0.25 < 0.2 & NO \\
\hline
Overall Accuracy Equality & 0.02 < 0.2 & YES \\
\hline
Mean Difference & 0.26 < 0.2 & NO \\
\hline
\makecell[l]{Equalized Odds (true positive) \\ Equal Opportunity} & 0.23 < 0.2 & NO \\
\hline
\makecell[l]{Equalized Odds (false positive) \\ Predictive Equality} & 0.22 < 0.2 & NO \\
\hline
\makecell[l]{Conditional Use Accuracy Equality (true positive) \\ Predictive Parity} & 0.06 < 0.2 & YES \\
\hline
Conditional Use Accuracy Equality (true negative) & 0.06 < 0.2 & YES \\
\hline
Equal Calibration & 0.03 < 0.2 & YES \\
\hline
Positive Balance & 1.6 & ? \\
\hline
Negative Balance & 1.4 & ? \\
\hline
    \end{tabular}
    \caption{Fairness measures of unprivileged group being African-American.}
    \label{tab:results}
\end{table}

\subsection{Fairness Analysis of African-American group}
On first blush, it is curious that it satisfies disparate impact but not demographic parity. However, if we return to the definition, we would see that disparate impact is meant to be used when being marked positive is an advantaged thing, while here in the COMPAS example, being marked positive is a disadvantaged thing.

It deceives the eye due to its definition being a ratio that lacks symmetry upon interchange of its components. Henceforth we will exclude disparate impact from our analysis of COMPAS. On the contrary, the other measures remain unaffected since their definitions are the absolute values of a difference.

We can then immediately tell from the failing demographic parity, mean difference, and conditional statistical parity that COMPAS prediction results were unfair against African-American, even if we only consider the ones with prior crimes.

From the low overall accuracy equality and both conditional use accuracy equality criteria, we can tell that the accuracy is similar across African-Americans and non-African-Americans.

From the failing equalized odds we can conclude that African-Americans are indeed treated unequally by the COMPAS system even after the ground truth is taken into account. This is the same conclusion reached by the ProPublica report.

From the low equal calibration we can tell that if we only consider the medium risk score, African-Americans are treated fairly.

Finally, if we look at positive and negative balance, we can see that they're of similar numbers. Their average is 1.5 on a scale of 10, which means approximately a 15\% difference. It remains to be said if this is fair or not. An auditor could consult a domain expert for advice on how to set a $\epsilon$ and how it is justified.

\subsection{Fairness Analysis of Different Races}
\begin{figure*}[h]
    \centering
    \begin{minipage}[b]{0.48\linewidth}
        \centering
        \includegraphics[width=\linewidth]{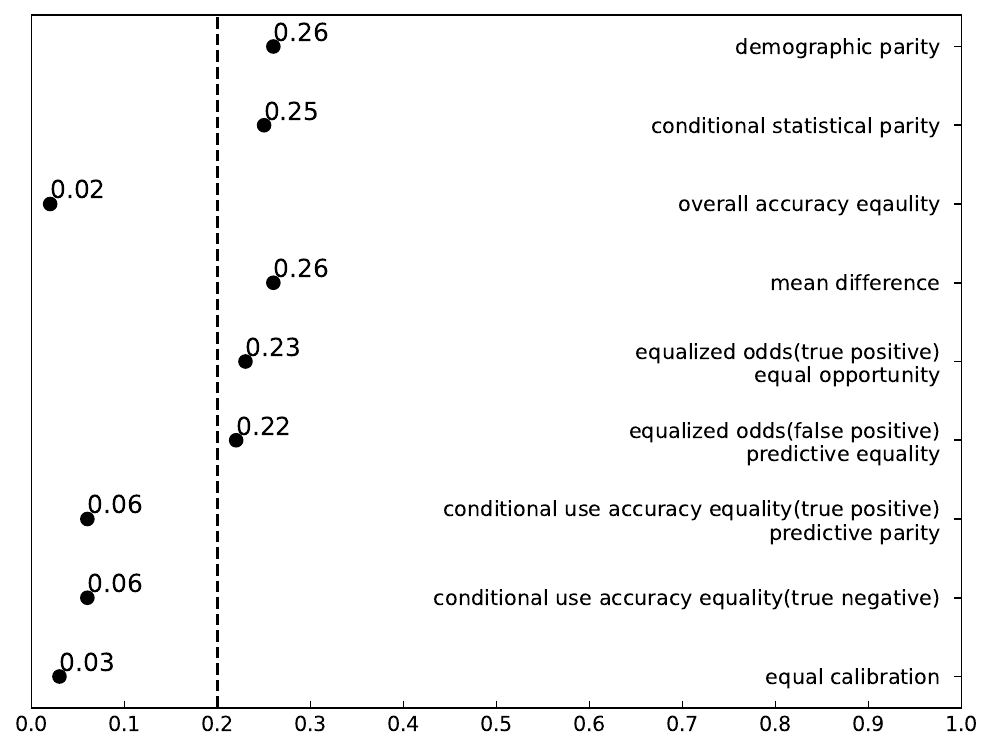}
        \subcaption{Unprivileged Group: African-American}\label{fig:1}
    \end{minipage}
    \hfill
    \begin{minipage}[b]{0.48\linewidth}
        \centering
        \includegraphics[width=\linewidth]{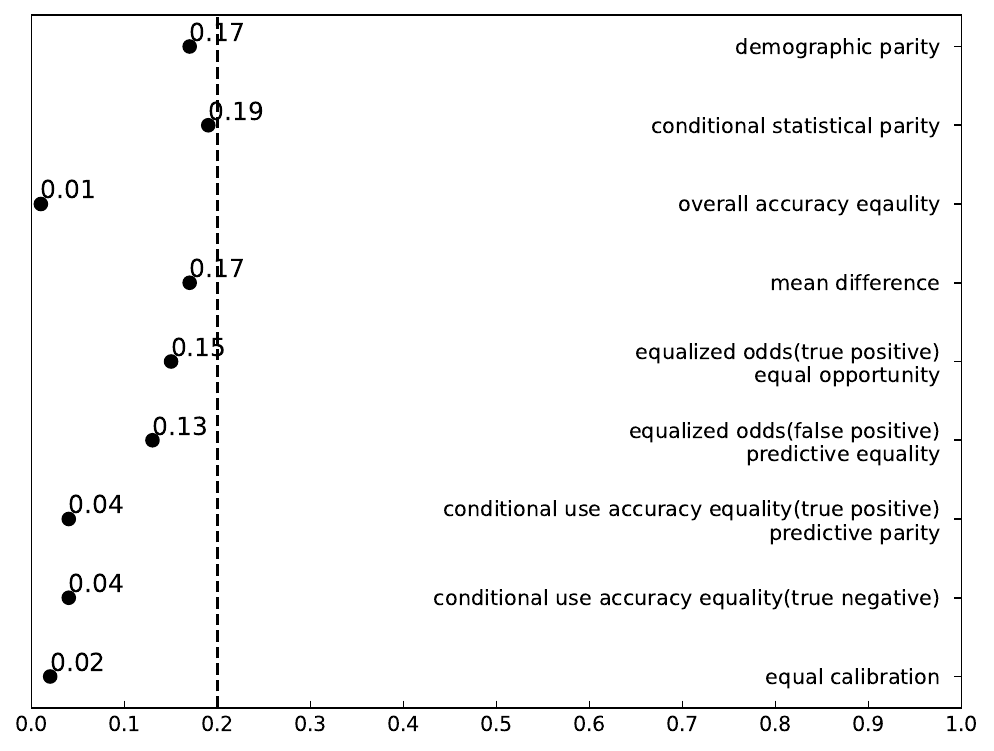}
        \subcaption{Unprivileged Group: Caucasian}\label{fig:2}
    \end{minipage}
    \hfill
    \begin{minipage}[b]{0.48\linewidth}
        \centering
        \includegraphics[width=\linewidth]{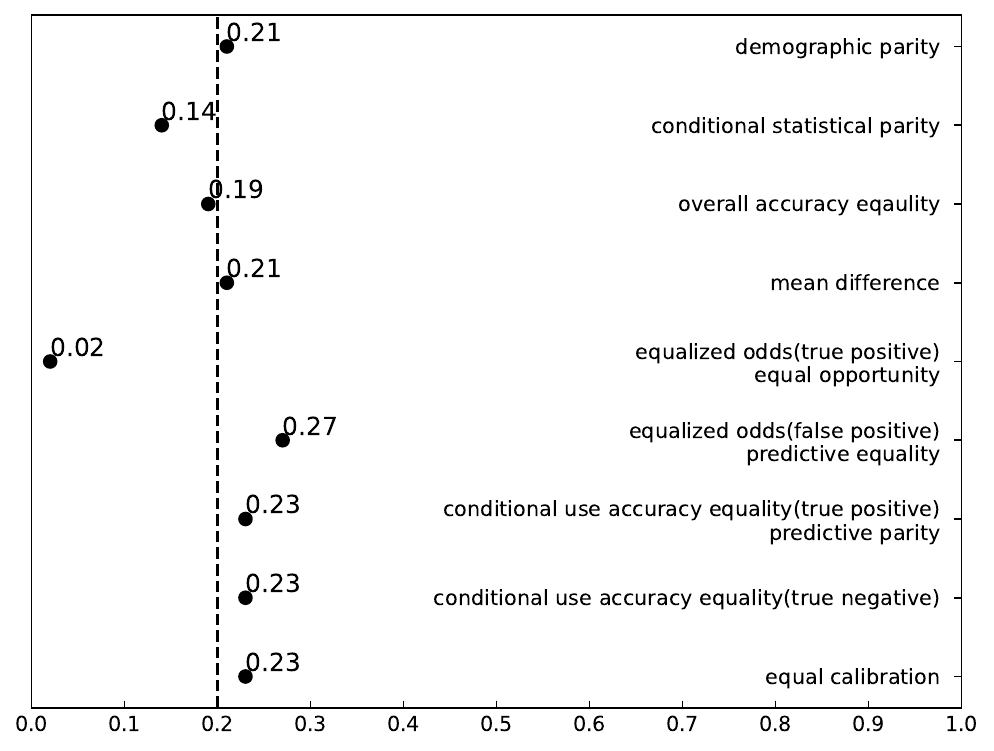}
        \subcaption{Unprivileged Group: Asian}\label{fig:2}
    \end{minipage}
    \hfill
    \begin{minipage}[b]{0.48\linewidth}
        \centering
        \includegraphics[width=\linewidth]{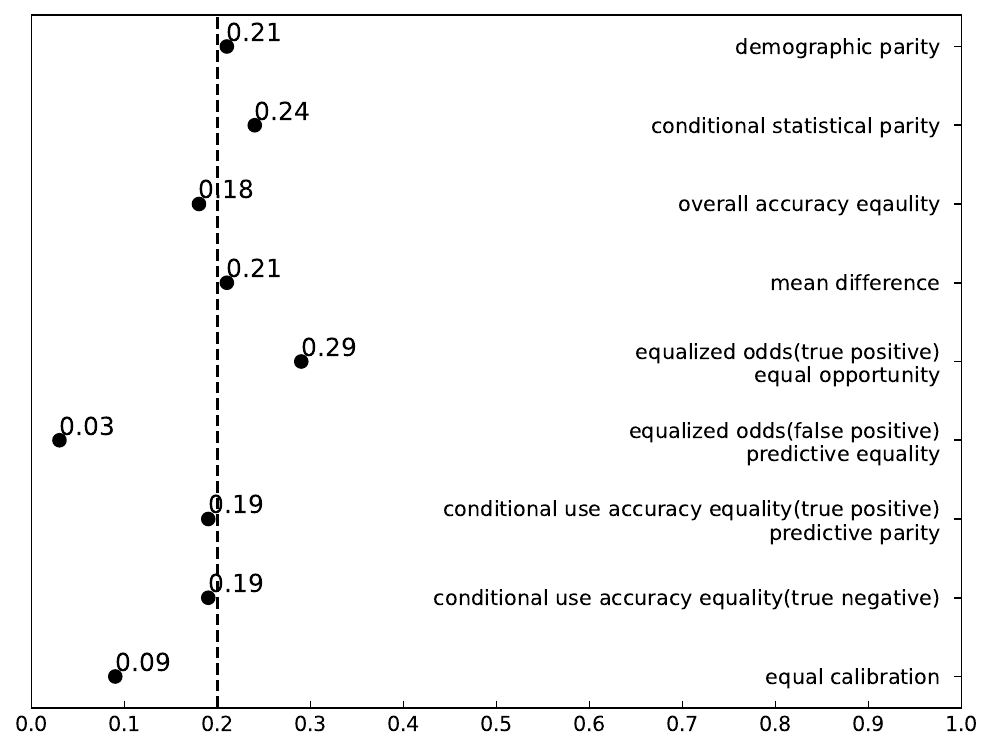}
        \subcaption{Unprivileged Group: Native American}\label{fig:2}
    \end{minipage}
    \hfill
    \begin{minipage}[b]{0.48\linewidth}
        \centering
        \includegraphics[width=\linewidth]{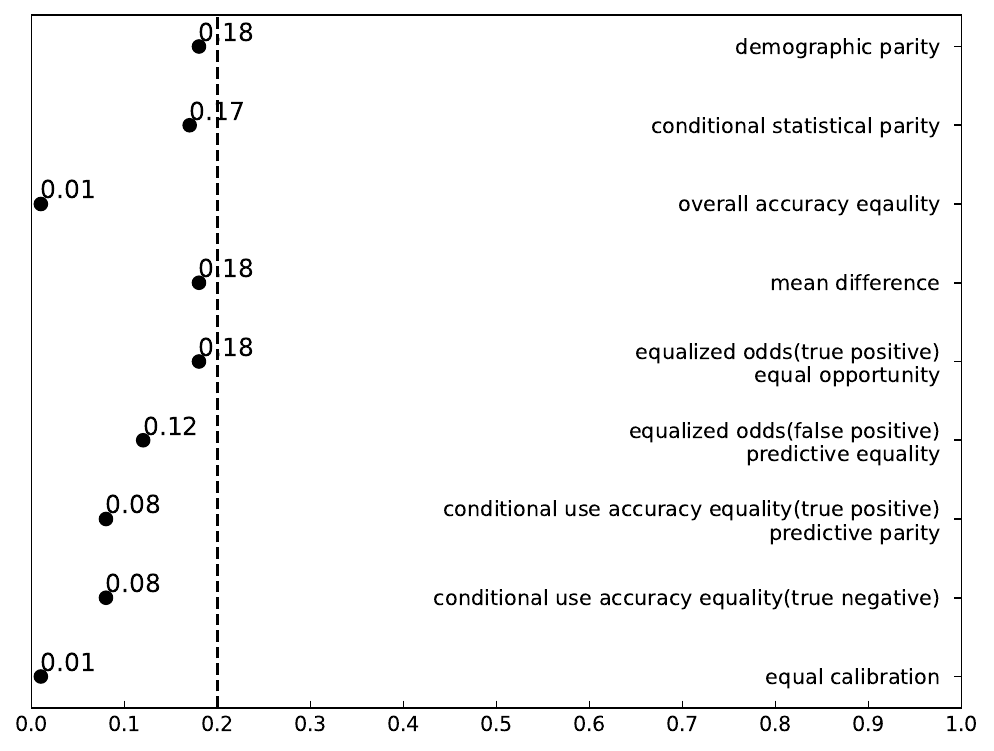}
        \subcaption{Unprivileged Group: Hispanic}\label{fig:2}
    \end{minipage}
    \hfill
    \begin{minipage}[b]{0.48\linewidth}
        \centering
        \includegraphics[width=\linewidth]{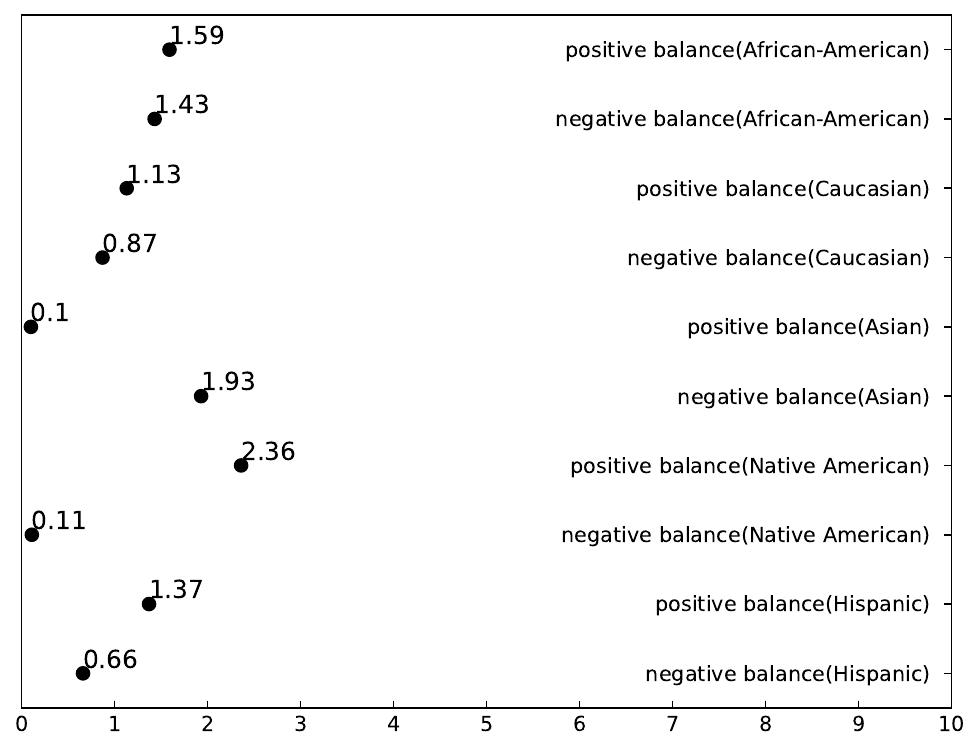}
        \subcaption{Balance}\label{fig:2}
    \end{minipage}
    \caption{Fairness measures of ProPublica COMPAS dataset grouped by race}
    \label{fig:compas}
\end{figure*}

By setting the privilege predicate to different races we can have a more comprehensive look over the dataset. We have checked the case of unprivilege predicate being African-American, Asian, Caucasian, Hispanic, and Native American. The results are shown in Figure~\ref{fig:compas}.

From the results, we can first notice that the overall accuracy equality is low for both the Asian and Native American cases. This can be explained by checking the original dataset which shows that there are only 32 and 18 rows, respectively, in a dataset of 7214 rows. Hence, the accuracy is naturally lower because of the small data size.

Looking at the remaining three groups, African-American, Caucasian, and Hispanic, their accuracies are relatively much better.

As for fairness measures, we can immediately see from that all measures are $\epsilon < 0.2$ for Caucasian and Hispanic that African-American are indeed treated unfairly in a broad sense.

More specifically, African-American are treated more unfairly according to demographic parity, conditional statistical parity, mean difference, and equalized odds. This is again in accordance with the conclusion of the ProPublica report.

\subsection{Fairness Analysis of Different Groups}
\begin{figure*}[h]
    \centering
    \begin{minipage}[b]{0.48\linewidth}
        \centering
        \includegraphics[width=\linewidth]{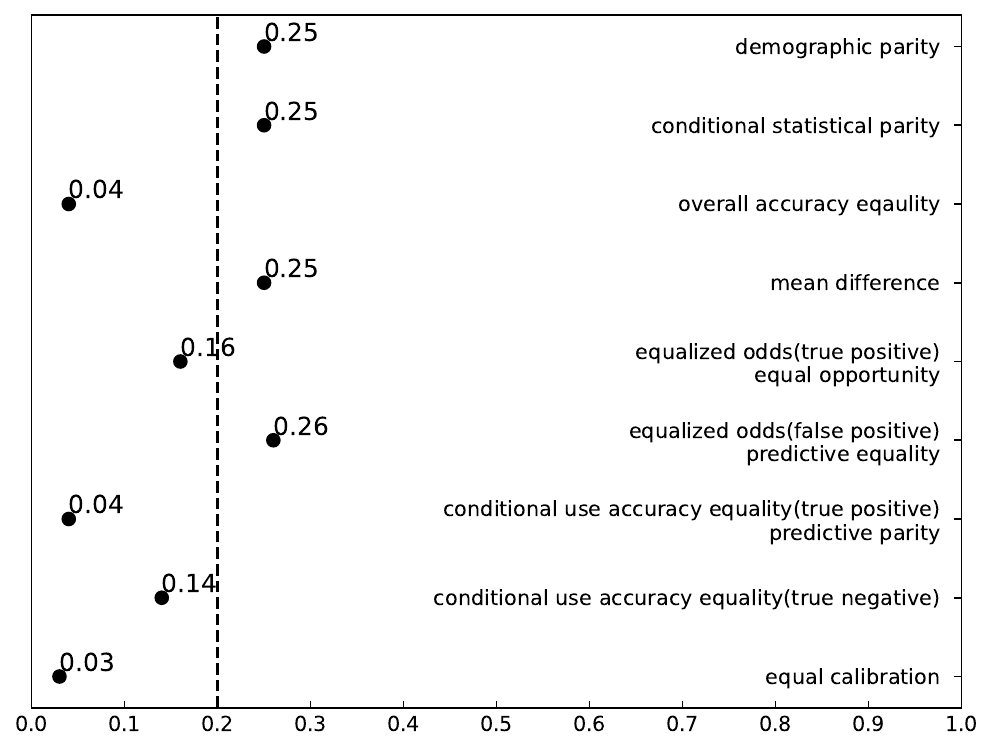}
        \subcaption{Unprivileged Group: Less than 25}\label{fig:1}
    \end{minipage}
    \hfill
    \begin{minipage}[b]{0.48\linewidth}
        \centering
        \includegraphics[width=\linewidth]{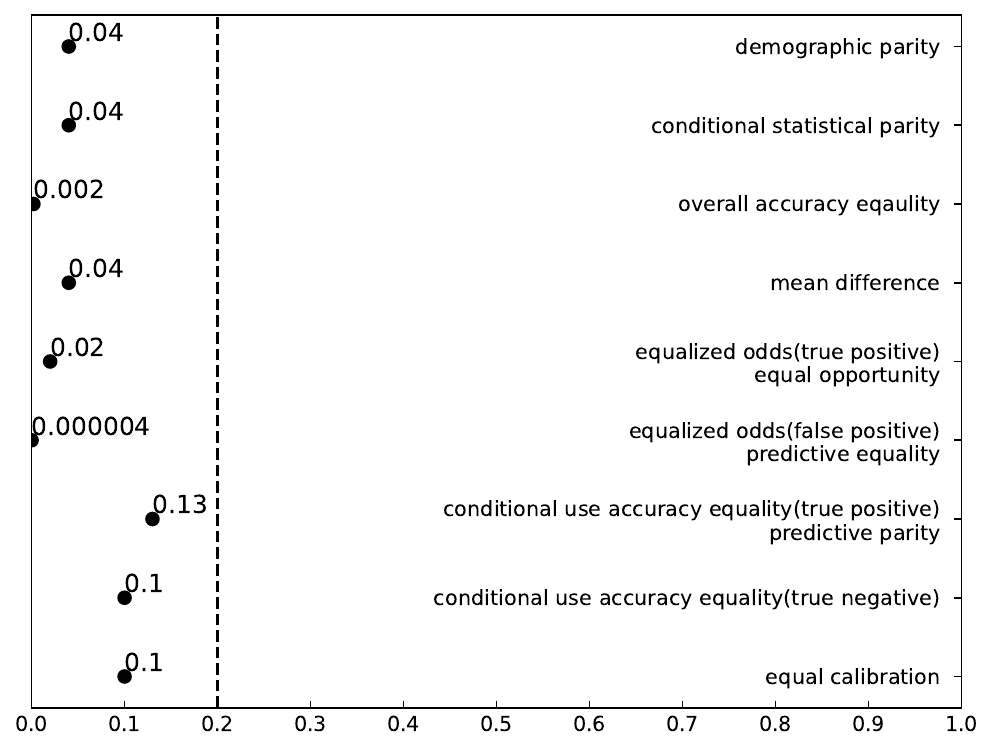}
        \subcaption{Privileged Group: Male}\label{fig:2}
    \end{minipage}
    \hfill
    \begin{minipage}[b]{0.48\linewidth}
        \centering
        \includegraphics[width=\linewidth]{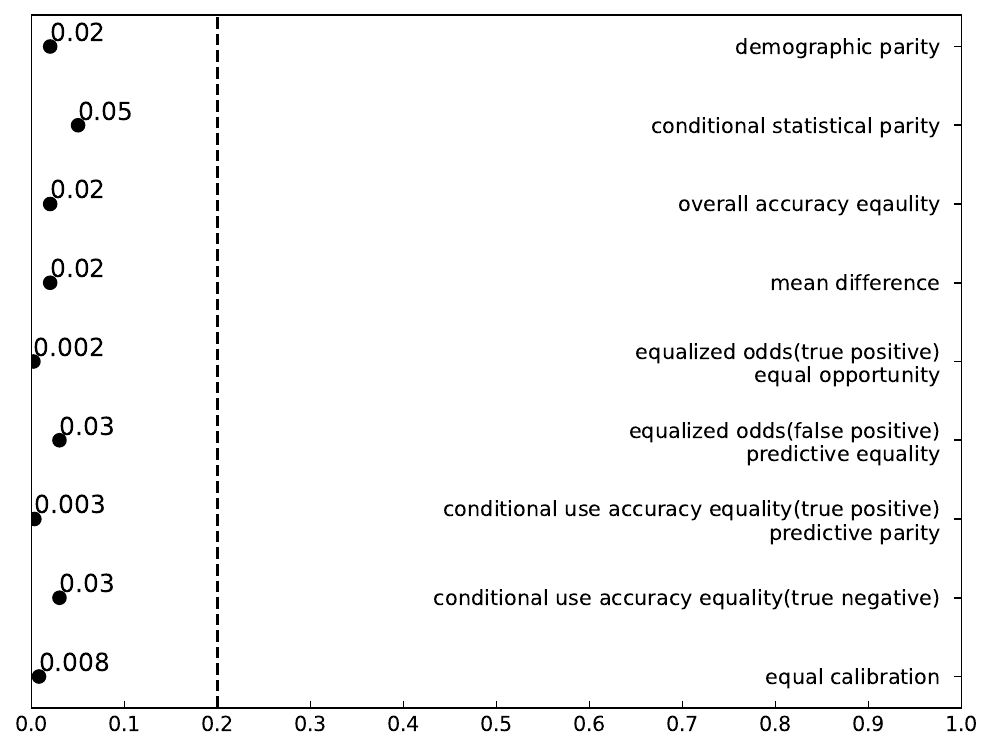}
        \subcaption{Unprivileged Group: 25 - 45}\label{fig:2}
    \end{minipage}
    \hfill
    \begin{minipage}[b]{0.48\linewidth}
        \centering
        \includegraphics[width=\linewidth]{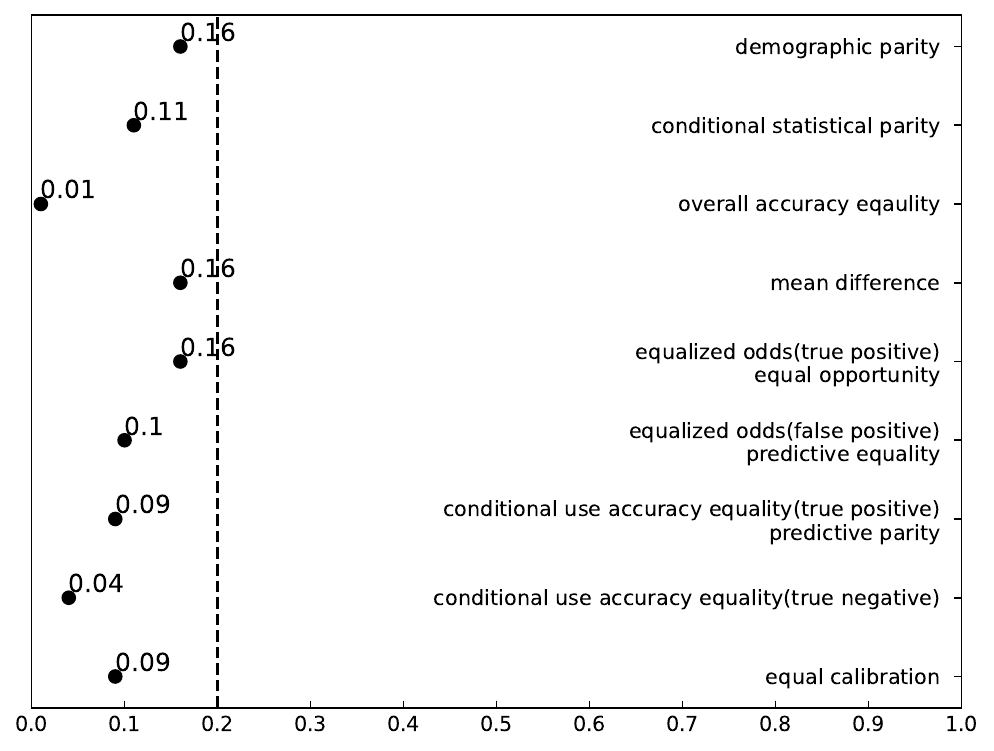}
        \subcaption{Privileged Group: Misdemeanor}\label{fig:2}
    \end{minipage}
    \hfill
    \begin{minipage}[b]{0.48\linewidth}
        \centering
        \includegraphics[width=\linewidth]{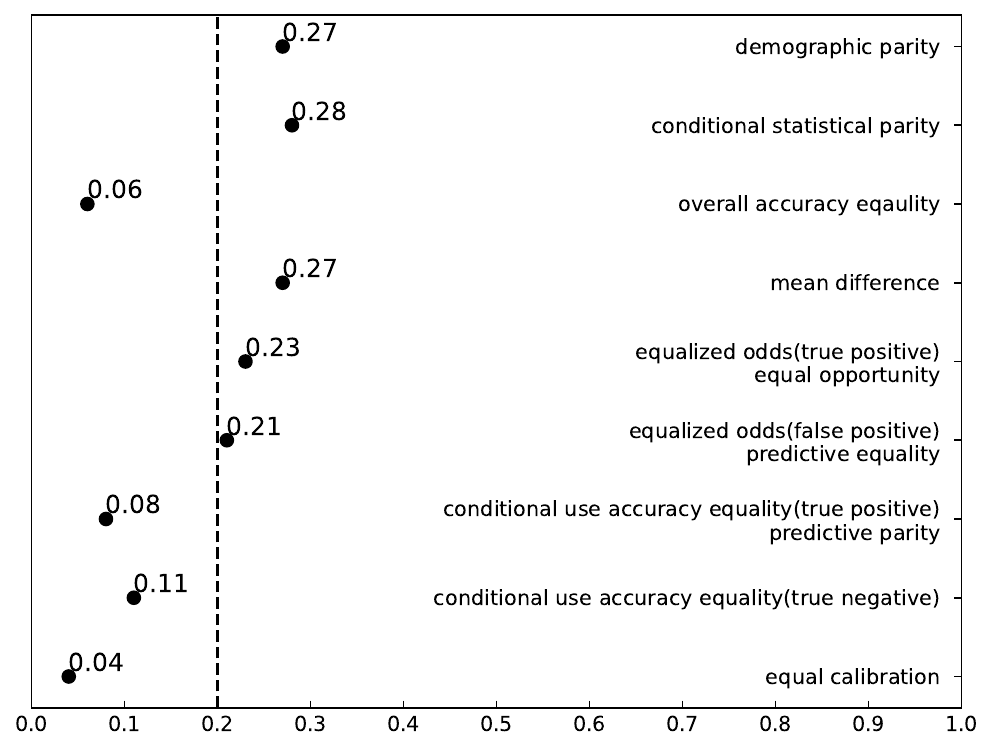}
        \subcaption{Unprivileged Group: Greater than 45}\label{fig:2}
    \end{minipage}
    \hfill
    \begin{minipage}[b]{0.48\linewidth}
        \centering
        \includegraphics[width=\linewidth]{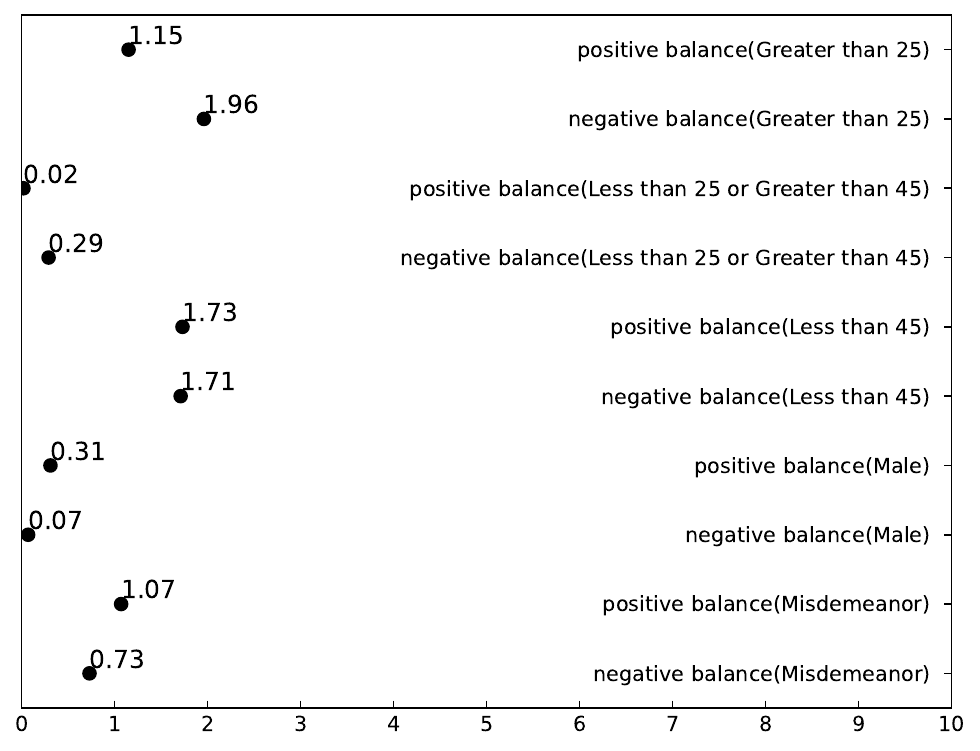}
        \subcaption{Balance}\label{fig:2}
    \end{minipage}
    \caption{Fairness measures of ProPublica COMPAS dataset grouped by age, sex, and charged degree}
    \label{fig:compas2}
\end{figure*}

On the other hand, we also analyzed the case of unprivileged group being across the three age groups and the case of privileged group of sex being Male and charged degree being misdemeanor in Figure~\ref{fig:compas2}.

We can see that, interestingly, the group of age ``25 - 45'' group receives generally fair treatment. Its data size is also the largest at 4109 rows, whereas ``Less than 25'' age group has 1529 rows and ``Greater than 45'' age group has 1576 rows, so the possibility of skewed data is unlikely.

If we look closer we can see that in the age group ``Less than 25'', its predictive equality is noticeably worse, meaning in young people false positive rate is higher than in older people; conversely, when we look at the age group ``Greater than 45'', both equal opportunity and predictive equality are worse, meaning for old people their treatment is even more unfair than young people. Only the middle age group is treated fairly.

In another vein, the case of ``Male'' group and ``Misdemeanor'' group are both treated rather fairly. Furthermore, their overall accuracy equality are both excellent at the $< 0.1$ range. The predictive equality  of ``Male'' group even has a measure as low as $0.000004$.

\subsection{Analysis with Model-as-input}
Although this dataset scenario falls in the CSV-as-input case in our framework, we also trained a simple makeshift model using the dataset itself for demonstration.

The model was trained by splitting the original dataset into 3 partitions: 60\% of the data was used for training; 20\% was used for validation; and the rest 20\% was used for calculating fairness measures. Validation showed that the accuracy was about 60\%.

We only chose equalized odds of the case where unprivileged group is African-American for illustrative purposes. The equalized odds of our model is 0.35 and 0.46, while that of the COMPAS dataset is 0.23 and 0.22. So, our simple naive model is more unfair than COMPAS's model and could use some more fine-tuning.

\section{Conclusion}
Addressing the challenges brought upon by the rapid advancements in AI technologies and its introduced bias necessitates a rigorous assessment of AI system fairness. While fairness is subjective, it can be defined through various statistical measures. Our research contributes to this effort by proposing a comprehensive framework for auditing AI systems using multiple white-box fairness metrics, such as demographic parity, equalized odds, and overall accuracy equality.

By applying this framework to the COMPAS dataset, we confirmed that African-American defendants were unfairly scored compared to other racial groups. These results align with ProPublica's findings, demonstrating the utility and accuracy of our fairness auditing tool. Developed in Python and publicly available, this tool enables third parties to conduct detailed assessments of AI systems, promoting transparency and accountability.

\bibliographystyle{IEEEtran}
\bibliography{references}

\end{document}